\documentclass[runningheads]{llncs}
\usepackage{graphicx}
\usepackage{soul}
\usepackage{booktabs}
\usepackage{amsmath}
\begin{document}
\title{Orientation-guided Graph Convolutional Network for Bone Surface Segmentation
}

\author{Aimon Rahman\textsuperscript{1}, 
Wele Gedara Chaminda Bandara\textsuperscript{1}, Jeya Maria Jose Valanarasu\textsuperscript{1}, Ilker Hacihaliloglu\textsuperscript{2}, Vishal M Patel\textsuperscript{1}} 
\institute{Johns Hopkins University \and University of British Columbia}

\maketitle              % typeset the header of the contribution
\begin{abstract}
Due to imaging artifacts and low signal-to-noise ratio in ultrasound images, automatic bone surface segmentation networks often produce fragmented predictions that can hinder the success of ultrasound-guided computer-assisted surgical procedures. Existing pixel-wise predictions often fail to capture the accurate topology of bone tissues due to a lack of supervision to enforce connectivity. In this work, we propose an orientation-guided graph convolutional network  to improve connectivity while segmenting the bone surface. We also propose an additional supervision on the orientation of the bone surface to further impose connectivity. We validated our approach on 1042 vivo US scans of femur, knee, spine, and distal radius. Our approach improves over the state-of-the-art methods by 5.01\% in connectivity metric.

\end{abstract}

\section{Introduction}

Computer assisted orthopedic surgery (CAOS) procedures use ultrasound (US) as it offers a cost effective and radiation free alternative to other modalities \cite{hacihaliloglu2017ultrasound}. Some non-surgical procedures such as \cite{yamauchi2009ultrasound,seitel2016ultrasound} also use US-based guidance systems. Accurate segmentation of bone surface from US images is essential for improved guidance in these procedures. Deep convolutional neural networks (CNNs) have been successfully adopted for this task while obtaining a decent performance. An UNet-like method was proposed in \cite{baka2017ultrasound} to localize vertebra bone surfaces. Methods like \cite{alsinan2019automatic,wang2018simultaneous,villa2018fcn} use a filtered feature-guided CNN for robust bone surface segmentation. Recently, a local phase tensor guided CNN is proposed in \cite{wang2020robust} and validated across US data of various characteristics.

 Although the above methods obtain decent segmentation performance with respect to measures like dice score, they do not have any specific measure to ensure bone connectivity. Note that the difference in the number of pixels between the joint and disjoint prediction is usually very less which results only a marginal difference in the dice score. Thus, the segmentation performance metrics do not quantify  bone connectivity. However, bone connectivity is essential as a disjoint  prediction can result in absurd clinical errors during CAOS procedures. Also,  bone surface segmentation maps are used in tasks like bone shadow segmentation. The bone shadow information is essential to guide the orthopedic surgeon to a standardized viewing plane with minimal noise and artifacts. If the bone segmentation map is disjoint, it can adversely affect the shadow predictions as it is very likely  for the shadow maps to become discontinuous. Hence, bone connectivity is an import factor to address in bone surface segmentation from US images.
 
 To this end, we propose an \textbf{O}rientation-guided \textbf{G}raph \textbf{C}onvolutional \textbf{N}etwork (\textbf{O-GCN}) for bone surface segmentation. First, we focus on improving connectivity by incorporating graph convolutions in the network architecture. Graph convolutional networks (GCN) are good at capturing relations among arbitrary regions in the input space than CNNs \cite{chen2019graph} which makes perfect sense for improving bone connectivity. We design a segmentation network in a UNet \cite{ronneberger2015u} like fashion but with GCNs instead of CNN blocks. Next, we propose utilizing orientation \cite{batra2019improved} as an additional supervision to obtain refined segmentation maps with improved connectivity. Learning orientation helps impose a connectivity constraint as learning the bone orientation favours connected bone surfaces. Also, supervising the network for both segmentation and orientation helps use get a more generalized features helping in better segmentation as well as improved connectivity. We validate the proposed O-GCN with 1024 vivo US scans of femur, knee, spine and distal radius and achieve better performance than recent methods.

In summary, the following are the contributions of this work: 

\begin{itemize}
    \item We propose a new network O-GCN for bone surface segmentation. O-GCN uses GCNs to model a better relationship focusing on improving connectivity and better segmentation. 
    
     \item We propose an orientation guided supervision for training O-GCN to help impose a connectivity constraint as learning the bone orientation favours connected bone surfaces.
    
    \item We conduct extensive experiments using the in vivo US scans of knee, femur, distal radius, spine, and tibia bones collected using two US machines and demonstrate that the proposed method is competitive with recent methods.
    
\end{itemize}

\section{Method}
\subsection{Graph convolution for bone segmentation}
The main building block of our O-GCN is Graph Convolution module which extracts the connectivity between bone segments by first constructing a graph $G = (\mathcal{V}, \mathcal{\xi}, A)$ in hidden feature domain, and then performing the graph convolution on the constructed graph G as: 
\begin{equation}
    \widetilde{X} = \sigma(A X W)
    \label{eq:graph_conv}
\end{equation}, where $\mathcal{V}$ denotes the nodes, $\mathcal{\xi}$ are the edges, $A$ is the similarity/adjacency matrix that describes the similarity between each node, $X$ is the input feature map, $W$ is the learnable weight matrix, $\sigma(\cdot)$ is the non-linear activation function, and $\widetilde{X}$ is the output from the graph convolution module. Different from the standard convolution operation $\widetilde{X} = \sigma(XW)$, the graph convolution incorporates the connectivity between the nodes to the convolution process through the similarity matrix $A$ which aids in improving bone segmentation prediction with less discontinuities.

In particular, given a $X$ of size $c \times h \times w$, where $c$ denotes the number of channels, $h$ and $w$ are the width and height of $X$, we construct a similarity matrix of size $l \times l$, where $l=h \times w$. Although there are various ways to compute the similarity between two nodes such as Euclidean distance and dot product similarity, we experimentally found that they introduce massive computational overhead and longer inference time making it unsuitable for medical image segmentation. Therefore, to tackle this problem, we propose a learnable similarity matrix $A(X)$, where the parameters of $A$ are adjusted during the learning process which ultimately resulting in less-memory requirement and faster inference time. Concretely, we compute the learnable similarity matrix $A(X)$ as follows: 
\begin{equation}
    A(X) = \sigma(\phi(X) \Lambda(X) \phi(X)^T), 
    \label{eqn:gcn}
\end{equation} 
where $\phi(X) \in \mathcal{R}^{l \times m}$ is modeled as $1\times1$ convolutional layer followed by ReLU, $\Lambda(X) \in \mathcal{R}^{m \times m}$ is a diagonal matrix by Global Average Pooling (GAP) followed by $1\times1$ convolution, and $T$ denotes the matrix transpose operation. After that, we compute the output from the graph convolution module  according to the eqn. (\ref{eq:graph_conv}), where $W$ is modeled as a $1\times1$ convolutional layer.

\subsection{Bone orientation learning as an auxiliary task}
\begin{figure}[tbh]
    \centering
    \includegraphics[width=\linewidth]{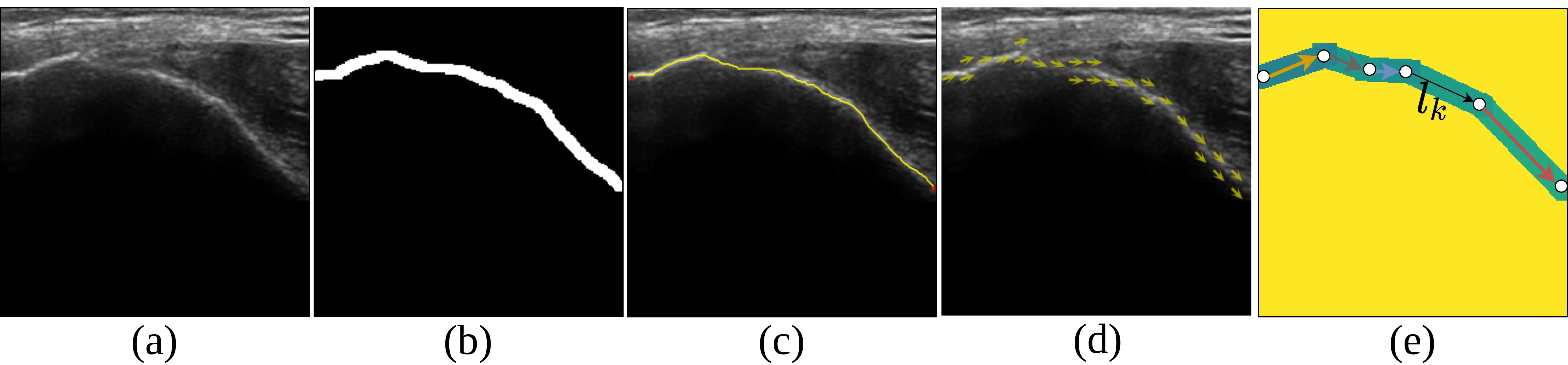}
    \caption{Generating bone orientation ground-truth from bone segmentation ground-truth. (a) US image. (b) Ground-truth bone segmentation mask. (c) Skeletonized ground-truth bone segmentation map. (d) Ground-truth bone orientation vectors $\vec{v}$. (e) Generated bone orientation ground-truth.}
    \label{fig:orient}
\end{figure}
Existing bone segmentation networks only utilize Binary Cross-Entropy (BCE) loss to optimize the network parameters. However, we argue that it fails to predict topologically correct and connected bone segments due to the absence of connectivity supervision in the training process. Motivated by this and the human behavior of annotating bones by tracing them in a specific orientation, we formulate an auxiliary bone orientation prediction task to enforce connectivity constraint into the main bone segmentation task.
%Orientation learning has been utilized in road segmentation problems to impose additional connectivity constraints \cite{bandara2021spin} \cite{batra2019improved}. This comes from the intuition that, the joint learning of related tasks will improve the encoded representation and increase the generalizability of the network. Utilizing orientation as an additional supervision signal to improve segmentation performance is inspired by the human tendency to annotate in a specific direction.

To this end, we employ the following procedure to generate the bone orientation maps from the ground-truth bone segmentation maps, that will be further used to calculate auxiliary bone orientation loss during the training process. Given a ground-truth bone segmentation mask (see Fig. \ref{fig:orient}-(b)), we first obtain bone line strings by skeletonizing the ground-truth bone segmentation map and smoothing it using RamerDouglas-Peucker algorithm~\cite{skel1,skel2} (see Fig. \ref{fig:orient}-(c)). Let's denote the set of bone line strings as $\{l_1, l_2, \cdots, l_k, \cdots, l_n\}$ and the 2D points connecting a given line string $l_k$ as $\{p_1, p_2, \cdots, p_n\}$. Assuming the surface segmentation is undirected, the points are sorted such that the vectors point from left to right and top to bottom to enforce a neural network to learn the connected representation (see Fig. \ref{fig:orient}-(c)). Between two consecutive coordinates pairs $(p_1, p_2), \cdots ,(p_{n-1}, p_{n})$ of $s_k$ we calculate a unit directional vector $\vec{v}$ as:
\begin{equation}
    \vec{v}_{i j}(x, y) =\frac{p_{i}(x, y)-p_{j}(x, y)}{\left\|p_{i}(x, y)-p_{j}(x, y)\right\|_{2}^{2}},
    \label{eq:vec_dir}
\end{equation}
where $(x, y)$ are the coordinates of points. Next, we obtain the orientation angle $o_r$ by converting it in to the polar domain as follows:
\begin{equation}
    \vec{v}_{i j}(x, y) \equiv\left\langle\begin{array}{ll}
1, & \left.\angle o_{r}\right\rangle
\end{array}\right.
\label{eq:angle}
\end{equation}
Finally, for each point pair $(p_i,p_j)$, the bone pixels lying within the threshold width of $\lambda_t$ along the perpendicular direction of $l_k$, we assign the same orientation value $o_r$, and for all the other none-bone pixels the non-bone orientation angle $o_{nb}$ is assigned as follows:
\begin{equation}
    o_{l_{k}}(m) = \begin{cases}o_{r} & \text { if }\left|\vec{v}_{\perp} \cdot \overrightarrow{\left(m-p_{1}\right)}\right|<\lambda_{t} \\
o_{nb} & \text { otherwise. }\end{cases},
\label{eq:angle_assign}
\end{equation}
where $o$ denotes the ground truth for orientation learning. In this work, we quantize ground-truth bone orientation angles $o$ into 26 bins; hence  formulating the auxiliary bone orientation leaning as pixel-wise multi-class classification problem. Fig \ref{fig:orient}-(e) shows how the final orientation ground-truth mask looks like for a given bone segmentation ground-truth mask.

\subsection{Network Details} 
  
\begin{figure}[!htb]
\centering
\includegraphics[width=1\linewidth]{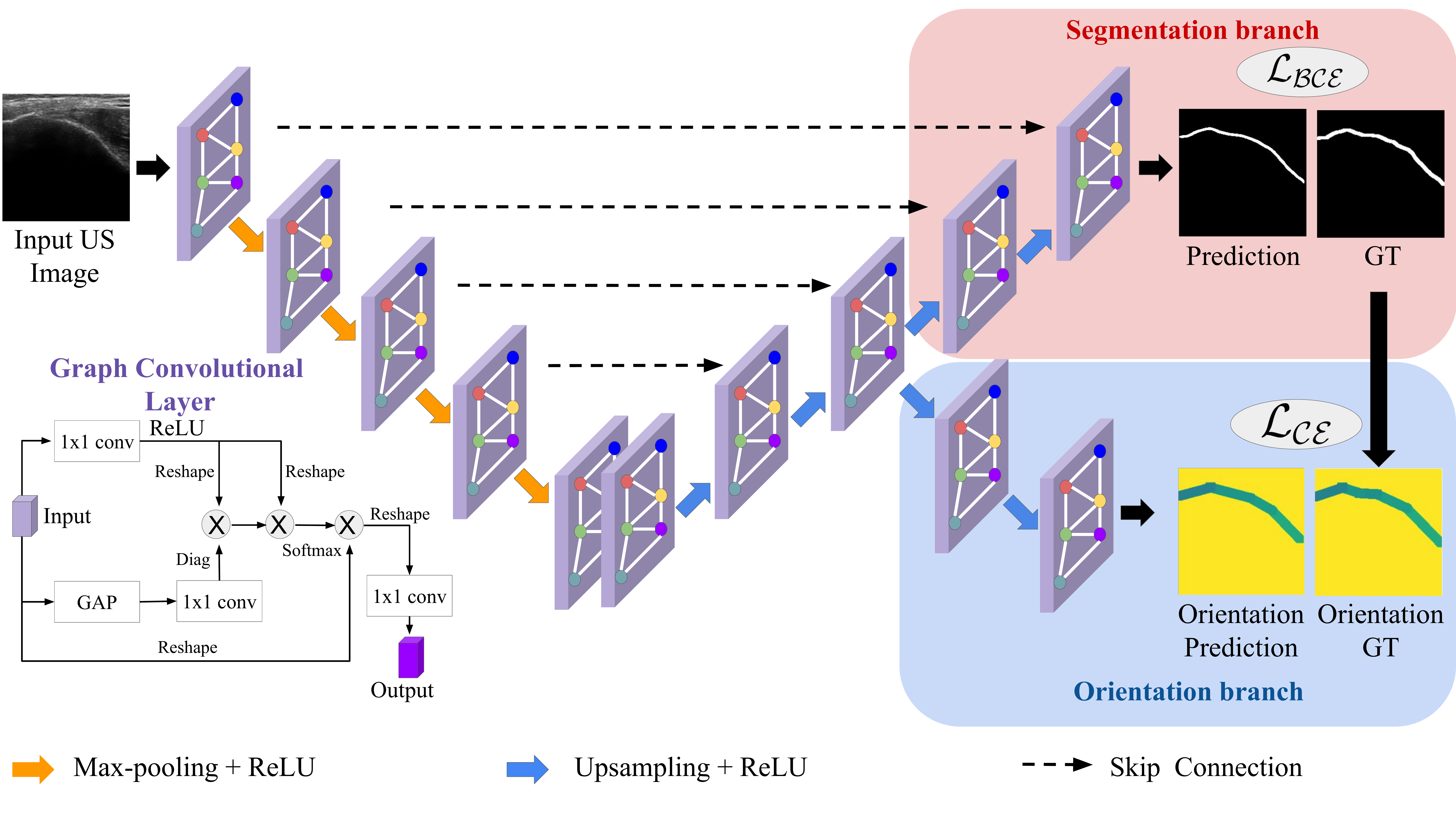}
\caption{An overview of the proposed O-GCN.
}
\label{fig:gcn_network}
\end{figure}

As shown in Fig. \ref{fig:gcn_network}, O-GCN follows a UNet \cite{ronneberger2015u} skeleton architecture but with GCNs instead of convolutional blocks. We use 5 GCN blocks in both the encoder and decoder. Each GCN block in the encoder is followed by a max-pooling layer and ReLU activation. Each GCN block in the decoder is followed by an upsampling layer and ReLU activation. For upsampling, we use bilinear interpolation. The output of the decoder is passed through a final $1 \times 1$ conv layer to get the segmentation map. For orientation learning, we create an auxiliary orientation branch from the 3-rd level of the decoder. Next, we stack two GCN modules with intermediate upsampling and ReLU non-linearity to bring the orientation prediction maps to the same spatial size as input, as shown in Fig. \ref{fig:gcn_network}. Note that the predicted orientation map $\hat{o}$  has 26 channels as we formulate orientation learning as a pixel-wise multi-class classification problem.

\section{Experiments and Results}
\subsection{Dataset}
The study is conducted on 1042 ultrasound images of the knee, femur, spine, and radius collected from 25 healthy volunteers with the approval of the institutional review board (IRB).  All data are collected using  SonixTouch US machine (Analogic Corporation, Peabody, MA, USA) with  2D C5-2/60 curvilinear and L14-5 linear transducer. A random split of 80:20  is applied for training and testing to the entire dataset based on subjects to avoid data leaking. The dataset is manually segmented by an expert ultrasonographer. 

\subsection{Implementation details} The network is trained using a batch size of 32 and trained until the convergence. Adam optimizer with a learning rate of $10^{-4}$ has been used to optimize the network parameters. They are updated by optimizing the following loss,
\begin{equation}
\mathcal{L}_{\text {seg }}(\hat{y}, y) =\operatorname{BCE}\left(\hat{y}, y\right) 
\end{equation}
\begin{equation}
 \mathcal{L}_{\text {orient }}(\hat{o}, o) =-\sum_{c=0}^{26} o_{c} \log \left(\hat{o}_{c}\right) 
\end{equation}
\begin{equation}
\mathcal{L}_{total} =\mathcal{L}_{\text {seg }}+0.5\mathcal{L}_{\text {orient }} 
\end{equation} where $o_c$ and $\hat{o}_c$ denote the $c^{th}$ class of $o$ and $\hat{o}$, respectively. All images are normalized between 0 to 1 and resized to 256 x 256 pixels before feeding to the network. All experiments are performed using a Linux workstation with Intel 3.50 GHz CPU and a 12GB NVidia Titan Xp GPU using the PyTorch framework.

\subsection{Evaluation Metrics} Although pixel-wise dice score is a commonly used metric for bone surface segmentation tasks, it is sub-optimal to evaluate the network’s ability to produce surface segmentation with accurate connectivity. In the context of bone surface segmentation, manual ground truths are not considered as the absolute gold standard, and detected true positives can be several millimeters away from the ground truth \cite{wang2018simultaneous}. Dice score will heavily penalize a slight error of bone surface width or its location however a briefly disconnected prediction will be penalized lightly. To this end, we utilize the Average Path Length Similarity (APLS) metric that has been widely used to measure connectivity like in road topography~\cite{etten_2017}. This measures the difference between ground truth and prediction by summing the differences in optimal path length between two nodes in the ground truth and prediction as follows:

\begin{equation}
APLS = 1-\frac{1}{N} \sum_i^N \min \left\{1, \frac{\left|L(a_i, b_i)-L\left(a_i^{\prime}, b_i^{\prime}\right)\right|}{L(a_i, b_i)}\right\}
\end{equation}
where $N$ is the number of unique bone surfaces in the ground truth bone segmentation map, and 
$L(a_i, b_i)=$ indicates the length between $path(a_i, b_i)$

\subsection{Quantitative Comparison} The quantitative results are presented in Table \ref{Tab:main}. Our proposed method achieves state-of-the-art results in terms of APLS and dice scores. In terms of APLS, the improvement is significant which demonstrates the effectiveness of the proposed network’s ability to predict bone surface with accurate connectivity.
\begin{table*}[!h]
\caption{Quantatitive comparison with the current state-of-the-art}
\label{Tab:main}
\centering
\medskip
\begin{tabular}{lcc}
\toprule
& \multicolumn{2}{c}{SonixTouch}\\
\cmidrule(lr){2-3}
Method & APLS ($\uparrow$) & Dice Score ($\uparrow$)\\
\midrule

Unet \cite{ronneberger2015u}  ~& $68.38\pm1.13$ & $80.22\pm0.13$  \\
MFG-CNN \cite{wang2018simultaneous} ~& $67.65\pm1.21$ & $82.18\pm0.21$ \\
LPT-GCT \cite{wang2020cnn}~ & $70.13\pm0.71$ & $82.91\pm0.16$ \\
Ours ~& $\mathbf{73.65\pm0.81}$& $\mathbf{83.19\pm0.22}$ \\

\bottomrule
\hline
\end{tabular}
\label{main}
\end{table*}

\subsection{Qualitative Results} 
\begin{figure}[!htb]
\centering
\includegraphics[width=1\linewidth]{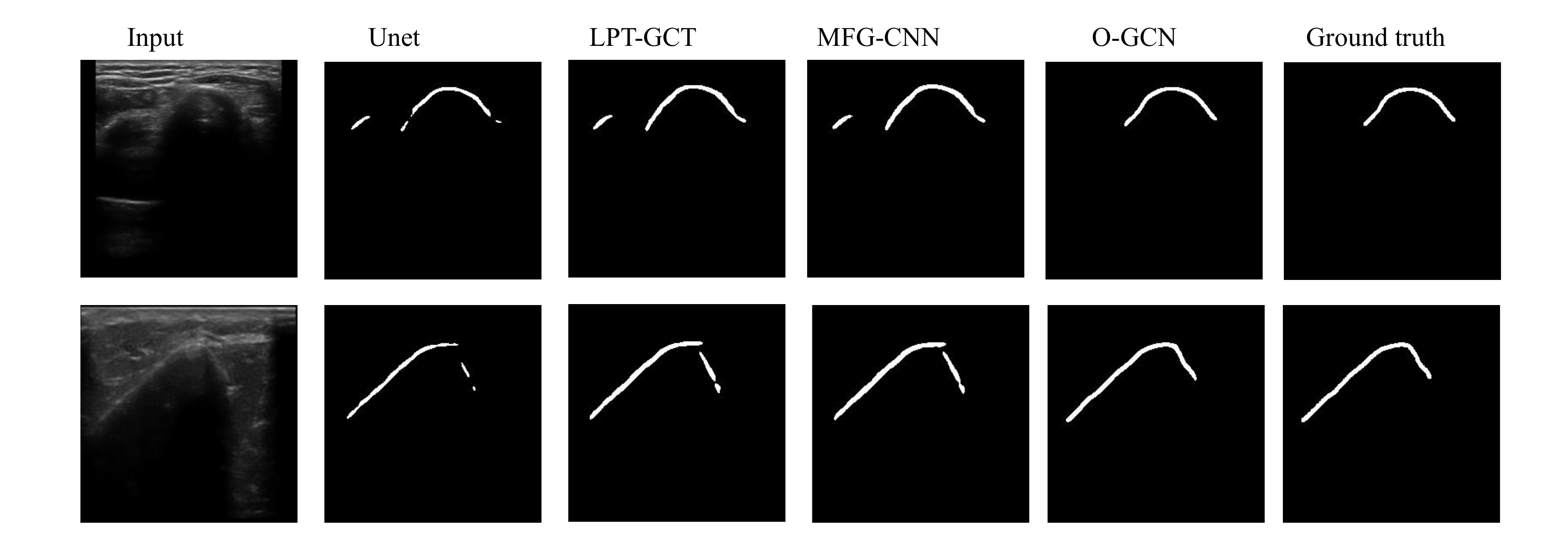}
\caption{A qualitative comparison with the state-of-the-art.}
\label{Fig:sota}
\end{figure}

For qualitative analysis, we visualize the predictions from Unet \cite{ronneberger2015u}, MFG-CNN \cite{wang2018simultaneous}, LPT-GCT \cite{wang2020cnn}, and our proposed method O-GCN in Fig. \ref{Fig:sota}. Due to low contrast in input ultrasounds,  baseline networks struggle to predict correct bone surfaces. In contrast, our method predicts bone surface without any discontinuity as the network captures long-range dependencies through graph-reasoning and utilizes orientation learning as indirect connectivity supervision.

\section{Discussion}

\subsection{Ablation Study} 
\begin{figure}[!htb]
\centering
\includegraphics[width=1\linewidth]{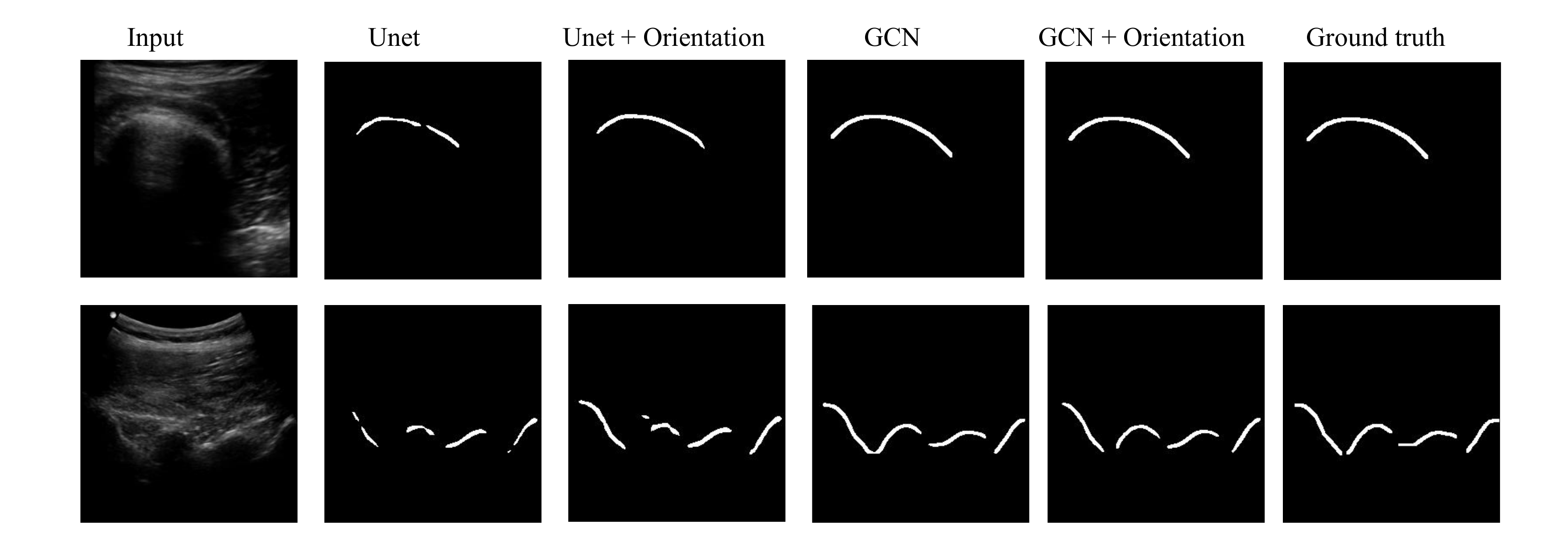}
\caption{A qualitative comparison for the ablation study.}
\label{Fig:abl}
\end{figure}

We perform an ablation study to prove the necessity of GCN-based architecture and orientation loss for bone surface segmentation. Table \ref{Tab:abl} shows that using GCN-based architecture boosts the performance by 4.40\%. Combining both GCN and orientation loss together results in further improvement in both metrics. Additionally, from the qualitative comparison in Fig. \ref{Fig:abl} it is clear that our method minimizes the fragmentation in bone surface segmentation.

\begin{table*}[!h]
\caption{Ablation Study}
\label{Tab:abl}
\centering
\medskip
\begin{tabular}{lcc}
\toprule
& \multicolumn{2}{c}{SonixTouch}\\
\cmidrule(lr){2-3}
Method & APLS ($\uparrow$) & Dice Score ($\uparrow$)\\
\midrule

ConvNet Only (Unet) \cite{ronneberger2015u}  ~& $68.38\pm1.13$ & $80.22\pm0.13$  \\
ConvNet + Orientation loss ~& $70.11\pm0.97$ & $81.95\pm0.31$ \\
GCN Only~ & $71.39\pm0.93$ & $82.77\pm0.24$ \\
GCN + Orientation loss ~& $\mathbf{73.65\pm0.81}$& $\mathbf{83.19\pm0.22}$ \\

\bottomrule
\hline
\end{tabular}
\label{main}
\end{table*}

\subsection{Bone Orientation loss} 	To study the performance of orientation learning, we choose both convnet and   GCN-based architectures. We modify these models by adding additional decoders for dual-task learning. The results in Table \ref{Tab:abl} show that incorporating orientation learning as additional supervision improves overall APLS and dice score for both CNN and GCN-based architecture, thus proving that learning two highly related tasks improves the shared encoded feature representation which leads to better performance.

\subsection{CNN vs. GCN} Qualitative results from Fig. \ref{Fig:abl} and quantitative results from Table \ref{Tab:abl} confirm the effectiveness of graph convolution neural network. CNN has limited capability of learning long-range spatial relations, which is demonstrated in Figure \ref{Fig:abl}. CNN-based models fail to accurately segment bone surfaces from ultrasound images. In contrast, graph modules accurately segment the entire bone surface even in low contrast images due to their ability to capture long-range dependencies between surface pixels through graph reasoning.

\section{Conclusion}

In this work, we have introduced a bone-orientation guided graph convolution network (O-GCN) for enhanced bone surface segmentation network. To the best of our knowledge, this is the first study to leverage bone orientation to constraint the model to produce unbroken surface predictions. We also demonstrate the problem with pixel-wise classification for the bone surface segmentation task and utilize APLS metrics to better evaluate the model. We propose using an orientation guided supervision for training O-GCN to help impose a connectivity constraint as learning the bone orientation favours connected bone surfaces.  Finally, we demonstrate that our multi-learning framework for related tasks improves the network prediction as the encoder learns better feature representation and obtains better performance than recent methods.

\bibliographystyle{splncs04}
\bibliography{references}

\begin{thebibliography}{10}
\providecommand{\url}[1]{\texttt{#1}}
\providecommand{\urlprefix}{URL }
\providecommand{\doi}[1]{https://doi.org/#1}

\bibitem{alsinan2019automatic}
Alsinan, A.Z., Patel, V.M., Hacihaliloglu, I.: Automatic segmentation of bone
  surfaces from ultrasound using a filter-layer-guided cnn. International
  journal of computer assisted radiology and surgery  \textbf{14}(5),  775--783
  (2019)

\bibitem{baka2017ultrasound}
Baka, N., Leenstra, S., van Walsum, T.: Ultrasound aided vertebral level
  localization for lumbar surgery. IEEE transactions on medical imaging
  \textbf{36}(10),  2138--2147 (2017)

\bibitem{batra2019improved}
Batra, A., Singh, S., Pang, G., Basu, S., Jawahar, C., Paluri, M.: Improved
  road connectivity by joint learning of orientation and segmentation. In:
  Proceedings of the IEEE/CVF Conference on Computer Vision and Pattern
  Recognition. pp. 10385--10393 (2019)

\bibitem{chen2019graph}
Chen, Y., Rohrbach, M., Yan, Z., Shuicheng, Y., Feng, J., Kalantidis, Y.:
  Graph-based global reasoning networks. In: Proceedings of the IEEE/CVF
  Conference on Computer Vision and Pattern Recognition. pp. 433--442 (2019)

\bibitem{skel2}
Douglas, D.H., Peucker, T.K.: Algorithms for the reduction of the number of
  points required to represent a digitized line or its caricature.
  Cartographica: the international journal for geographic information and
  geovisualization  \textbf{10}(2),  112--122 (1973)

\bibitem{etten_2017}
Etten, A.V.: Spacenet road detection and routing challenge part ii - apls
  implementation (Nov 2017),
  \url{https://medium.com/the-downlinq/spacenet-road-detection-and-routing-challenge-part-ii-apls-implementation-92acd86f4094}

\bibitem{hacihaliloglu2017ultrasound}
Hacihaliloglu, I.: Ultrasound imaging and segmentation of bone surfaces: A
  review. Technology  \textbf{5}(02),  74--80 (2017)

\bibitem{skel1}
Ramer, U.: An iterative procedure for the polygonal approximation of plane
  curves. Computer graphics and image processing  \textbf{1}(3),  244--256
  (1972)

\bibitem{ronneberger2015u}
Ronneberger, O., Fischer, P., Brox, T.: U-net: Convolutional networks for
  biomedical image segmentation. In: International Conference on Medical image
  computing and computer-assisted intervention. pp. 234--241. Springer (2015)

\bibitem{seitel2016ultrasound}
Seitel, A., Sojoudi, S., Osborn, J., Rasoulian, A., Nouranian, S., Lessoway,
  V.A., Rohling, R.N., Abolmaesumi, P.: Ultrasound-guided spine anesthesia:
  feasibility study of a guidance system. Ultrasound in Medicine \& Biology
  \textbf{42}(12),  3043--3049 (2016)

\bibitem{villa2018fcn}
Villa, M., Dardenne, G., Nasan, M., Letissier, H., Hamitouche, C., Stindel, E.:
  Fcn-based approach for the automatic segmentation of bone surfaces in
  ultrasound images. International journal of computer assisted radiology and
  surgery  \textbf{13}(11),  1707--1716 (2018)

\bibitem{wang2018simultaneous}
Wang, P., Patel, V.M., Hacihaliloglu, I.: Simultaneous segmentation and
  classification of bone surfaces from ultrasound using a multi-feature guided
  cnn. In: International conference on medical image computing and
  computer-assisted intervention. pp. 134--142. Springer (2018)

\bibitem{wang2020robust}
Wang, P., Vives, M., Patel, V.M., Hacihaliloglu, I.: Robust real-time bone
  surfaces segmentation from ultrasound using a local phase tensor-guided cnn.
  International Journal of Computer Assisted Radiology and Surgery
  \textbf{15}(7),  1127--1135 (2020)

\bibitem{wang2020cnn}
Wang, P., Vives, M., Patel, V.M., Hacihaliloglu, I.: Robust real-time bone
  surfaces segmentation from ultrasound using a local phase tensor-guided cnn.
  International Journal of Computer Assisted Radiology and Surgery
  \textbf{15},  1127--1135 (2020)

\bibitem{yamauchi2009ultrasound}
Yamauchi, M., Kawaguchi, R., Sugino, S., Yamakage, M., Honma, E., Namiki, A.:
  Ultrasound-aided unilateral epidural block for single lower-extremity pain.
  Journal of anesthesia  \textbf{23}(4),  605--608 (2009)

\end{thebibliography}

\end{document}